\documentclass[preprint2]{aastex}

\shorttitle{Photometry and LF of M15}
\shortauthors{Feuillet, Paust, \& Chaboyer}

\begin{document}

\title{$BVI$ Photometry and the Red Giant Branch Luminosity Function of M15}

\author{Diane K. Feuillet}
\affil{New Mexico State University}
\affil{Department of Astronomy \\ New Mexico State University \\ P.O. Box 30001, MSC 4500 \\ Las Cruces, NM 88003}
\email{feuilldk@nmsu.edu}
\and
\author{Nathaniel E. Q. Paust}
\affil{Whitman College}
\affil{Whitman College Hall of Science \\ 345 Boyer Ave \\ Walla Walla, WA 99362}
\email{paustne@whitman.edu}
\and
\author{Brian Chaboyer}
\affil{Dartmouth College}
\affil{Department of Physics and Astronomy \\Dartmouth College \\6127 Wilder Laboratory \\ Hanover, NH 03755}
\email{chaboyer@heather.dartmouth.edu}

\begin{abstract}
We present new $BVI$ photometry containing 40,000 stars of the Galactic globular cluster M15 (NGC 7078) covering a $25 \times 25$ arcminute area centered on the cluster with a magnitude range from the tip of the Red Giant Branch to three magnitudes below the main sequence turn-off. Using $\alpha$-enhanced Dartmouth Stellar Evolution Program models, we find an age of $13.0 \  \pm \ 1.0$ Gyr and distance modulus of $(m-M)_V = 15.4 \pm 0.1$ through isochrone fitting. Unlike previous works, we find good agreement between the observed completeness-corrected stellar luminosity function and models.
\end{abstract}


\section{Introduction}
Stellar evolution models are an essential tool in astronomical research.  At a basic level, isochrone fitting is the most commonly used method for stellar cluster age determination.  At higher levels, stellar evolution models are used to determine the stellar populations of distant galaxies from their integrated colors and the physical sizes of stars hosting extrasolar planets.  Particularly in the case of galaxies, it is important that the evolutionary timescales of evolved red giant branch (RGB) stars be well understood since these stars are often the brightest and most numerous stars.  Globular clusters (GCs), as large, single-age, single-metallicity populations, are the perfect test subject to determine accuracy of a stellar evolution model. While hundreds to thousands of papers utilize the color-magnitude diagram (CMD) of star clusters to determine age, metallicity, and the distance modulus, the luminosity function (LF) provides more information about stellar evolution both above and below the main-sequence turnoff (MSTO).  Below the MSTO, the LF reflects the initial mass function (IMF) possibly with dynamical mass segregation effects  and mass loss visible. Above the MSTO, the LF maps the speed and progress of the hydrogen-burning shell through the stars, revealing interesting features of stellar internal structure. 

\citet{bolte} and \citet{sandquist} conducted studies testing stellar evolution models with the luminosity functions (LFs) of M30 and M5 respectively. Those studies found that the models did not match the observations, prompting serious concerns about whether the internal structure of RGB stars was understood at low to intermediate metallicities. Extensive testing and research on stellar evolution has greatly improved the current models, such as the Dartmouth Stellar Evolution Program (DSEP) models, which do not suffer from the same LF disagreements \citep{paust}. However, ongoing testing is required to ensure the accuracy of these models and the LF of M15 (NCG 7078) is an excellent location to test these stellar evolution models.  

M15 has been well-studied due to an observationally advantaged position. M15 is a large cluster and at a sufficient distance from the galactic plane, with a $b = -27.31$ \citep{harris1996}, making study relatively easy without interference from external crowding, field star contamination, or dust.  For background, we present a small number of the works.  It has a low metallicity, [Fe/H] $= -2.26$ \citep{harris1996}, $-2.15$ \citep{zinn}, $-2.15$ \citep{carretta}, $-2.42$ \citep{kraft}, making it one of the lowest metallicity GGCs and a good test of extreme behavior.  \citet{corwin} identified variable stars in the cluster, resulting in an accurately determined distance modulus.  \citet{majaess} reexamined the distance modulus to several globular clusters, including M15, and found that they agreed with the literature. Using less precise data, they find a distance modulus of $(m-M) \ge 14.82$ which is consistent with the dereddened distance modulus from this work. \citet{mcnamara} conducted a study of M15 using proper motions and radial velocities to determine the cluster's age and mass, and the absolute magnitude of its RR Lyrae stars. They found an age of $13.2 \pm 1.5$ Gyr, which agrees with our age estimate. \citet{an} conducted a detailed test of theoretical stellar isochrones on several globular clusters, including M15, and concluded that the Yale Rotating Evolutionary Code models fit their data best.  \citet{ruelas} conducted a study of M15 similar to the study presented below. Our photometry extends to fainter magnitudes, including the full MSTO, and contains a larger RGB population. 

Previous studies of this cluster have revealed some interesting stellar dynamics in the cluster leading to the possibility of a central black hole \citep{pasquali, bash, murphy, bosch}.  \citet{pasquali} measured the cluster stellar mass function and examined how the luminosity function varied radially due to mass segregation in the $H$ and $J$ bands. They concluded that an intermediate mass central black hole was not necessary to explain the observed stellar dynamics \citep{pasquali}. \citet{bash} used the Very Large Array to examine three globular clusters, including M15, for radio emission and the possible presence of a central black hole. This study did not detect any radio emission suggestive of an intermediate mass black hole from the centers of any of the clusters studies, however, this data does not prove that globular clusters do not have central black holes. \citet{bosch} developed orbit-based dynamical models for M15 fit to radial velocities and proper motions. From this model, they were able to constrain the mass-to-light ratio $M/L$ as a function of radius. They confirmed a central mass concentration in the cluster, but were not able determine the exact nature of the concentration. Our research, however, focuses on the LF of the red giant branch (RGB), not the main sequence.  All of the evolved stars in the cluster have roughly the same mass, so mass segregation and stellar dynamics do not have a significant affect on our results. 

In this work, \S \ref{observations} details the observations of the cluster, while \S \ref{reductions} describes the reduction of the data. Section \ref{cmd} presents the data as CMDs and \S \ref{lfs} discusses the LFs, including determining completeness, generating the LFs, and fitting the data with DSEP models. We will focus on the fit of the models to the RGB in order to determine how well DSEP models stellar evolution. Conclusions about the models and LFs are discussed in \S \ref{conclusion}.

\section{Observations \label{observations}}

\begin{figure}
\epsscale{1.0}
\plotone{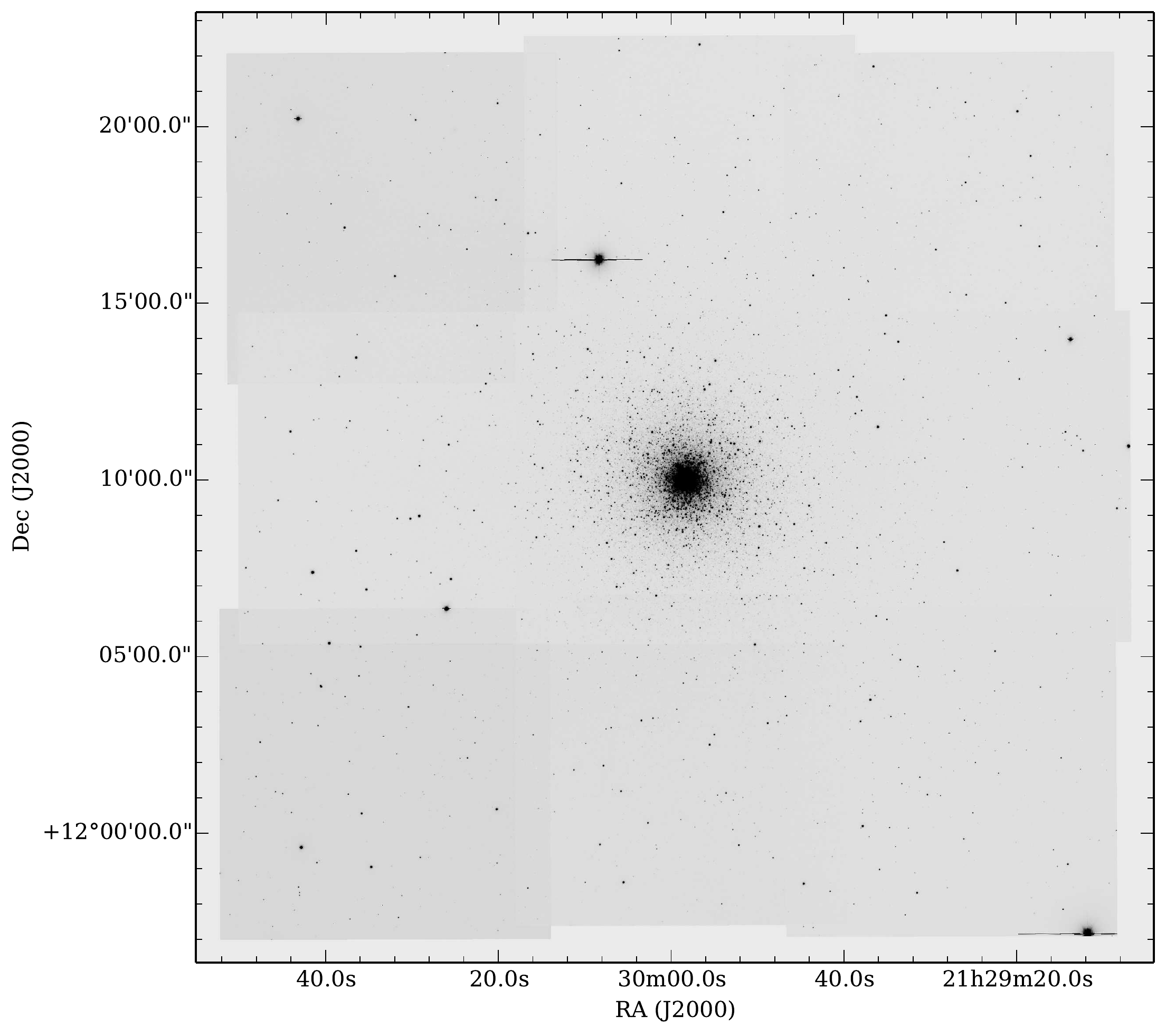}
\caption{M15  in the $V$ long exposure images.  Single images from each of the $3 \times 3$ grid positions were used to create this image of the $25' \times 25'$ area covered by the photometry in this work. 
\label{fig:M15image}}
\end{figure}

M15 was observed during three observing runs in November 2002, September 2003, and September 2004 using the Hiltner 2.4m telescope at the MDM Observatory at Kitt Peak, AZ.  All frames were taken using the Echelle camera, a $2048 \times 2048$ camera with a field of view of approximately $9\times9$ arcminutes. The images were tiled a grid to increase the areal coverage on the cluster.  Two separate grids were used, $3 \times 3$ for short exposures and $4 \times 4$ for long exposures, with the final full-depth photometry covering a $25' \times 25'$ area, as seen in Figure \ref{fig:M15image}.  The short exposures were approximately 30 seconds each and the long exposures were approximately 300 seconds each. Three or more dithered images were taken of each grid tile in each of the B, V, and I filters, for a total of over 200 images.  

\section{Data Reduction \label{reductions}}

\subsection{Image Processing and Photometry}

The images taken of M15 were reduced using standard IRAF techniques with nightly biases and twilight flats. Profile-fitting photometry was performed on the frames using the DAOPHOT and ALLSTAR programs developed by \citet{stetson1987,stetson1994}. Approximately 100 bright, uncrowded stars were selected in each image to characterize the point-spread function (PSF) and its variation about the frame. While the PSF stars selected were outside the most crowded regions of each frame, they sampled the full x- and y-range of each image to accurately map the PSF variations. 

The final photometry files were generated from the raw image photometry files in a multi-step process.  First, we filtered the raw photometry to remove any detection with a measured Poisson uncertainty greater than 0.1 mag. Second, applied an aperture correction derived from the brightest uncrowded stars from each frame.  The derived aperture correction was found to be independent of position.  Third, DAOMASTER and DAOMATCH were used to match the individual photometry files and combine them into one master file for each filter, requiring each star be detected at least twice.  The B, V, and I master files were filtered to contain only stars whose frame-to-frame magnitude variation was less  than 0.1 magnitude. Finally, the master files were then matched using DAOMASTER and DAOMATCH to make a composite file, requiring a star be detected in all three filters to be included in the final catalog.

\subsection{Calibration}

The data was calibrated by direct comparison to the \citet{stetson2009} photometric standards for M15. Comparing stars brighter than $V=17$, we identified over 200 stars from our photometry with established magnitudes. The best photometric solution to convert the data to the Stetson system was found to depend on color to the first order. The transformation equations are as follows.
\begin{eqnarray}
\nonumber B=b_\mathrm{inst}+2.321+0.024(b_\mathrm{inst}-i_\mathrm{inst})
\\
\nonumber V=v_\mathrm{inst}+4.171-0.026(b_\mathrm{inst}-i_\mathrm{inst})
\\
\nonumber I=i_\mathrm{inst}+0.981+0.011(v_\mathrm{inst}-i_\mathrm{inst})
\end{eqnarray}

\begin{figure}
\epsscale{1.0}
\plotone{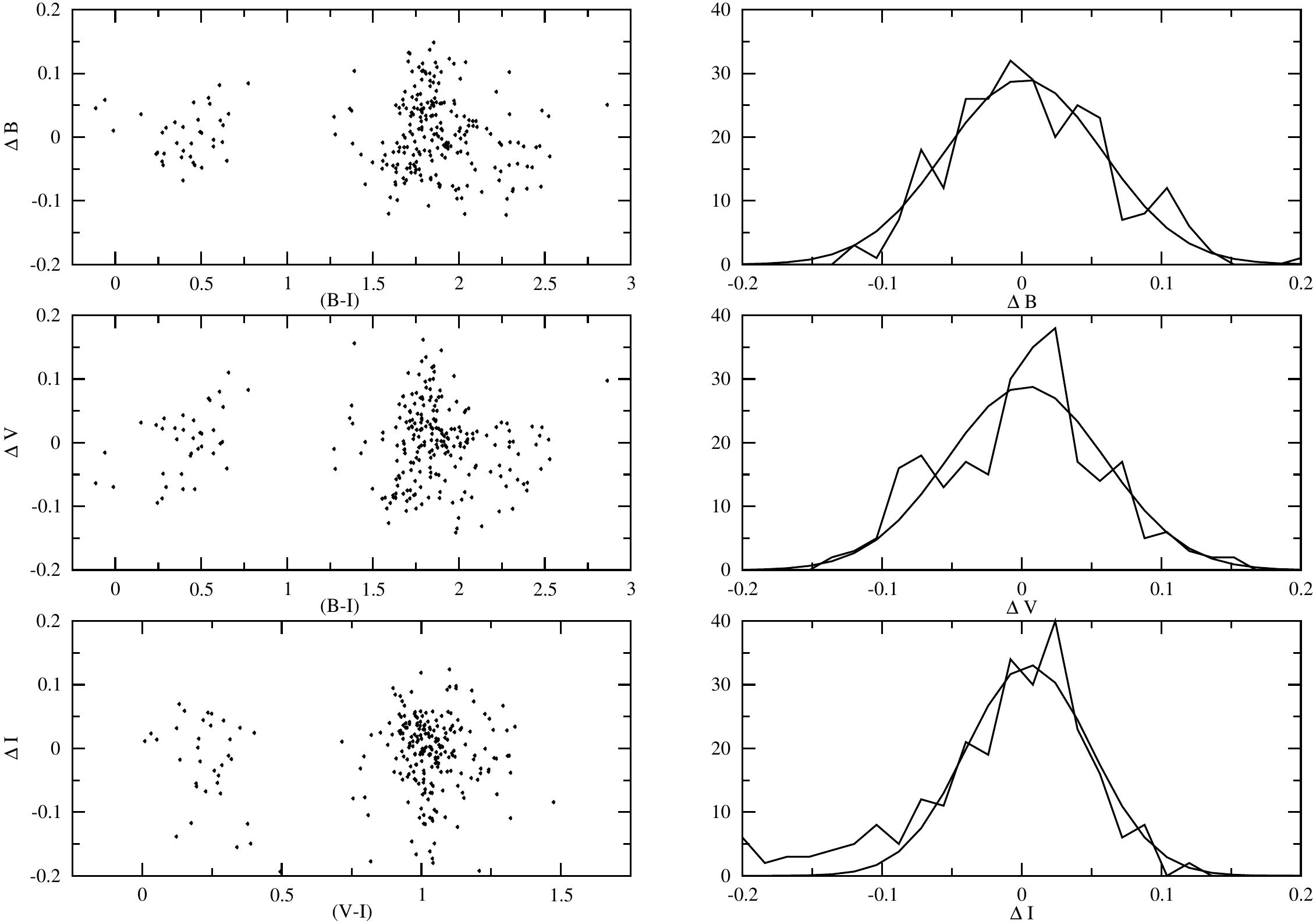}
\caption[Residual Distribution for Standard Stars]{The residual distributions for the standard stars for each filter. The second column of plots shows the error distribution of each color and the best-fit Gaussian. The standard deviations for the Gaussians are 0.057, 0.056, and 0.045 for $B$, $V$, and $I$ respectively. This matches the photometric errors.
\label{fig:colorcalib}}
\end{figure}

The residuals from the fit between the observed stars and the Stetson standards are shown in Figure \ref{fig:colorcalib}. In all cases, the distribution of the residuals matched the photometric errors.

\section{Color-Magnitude Diagrams \label{cmd}}

\begin{figure}
\epsscale{0.7}
\plotone{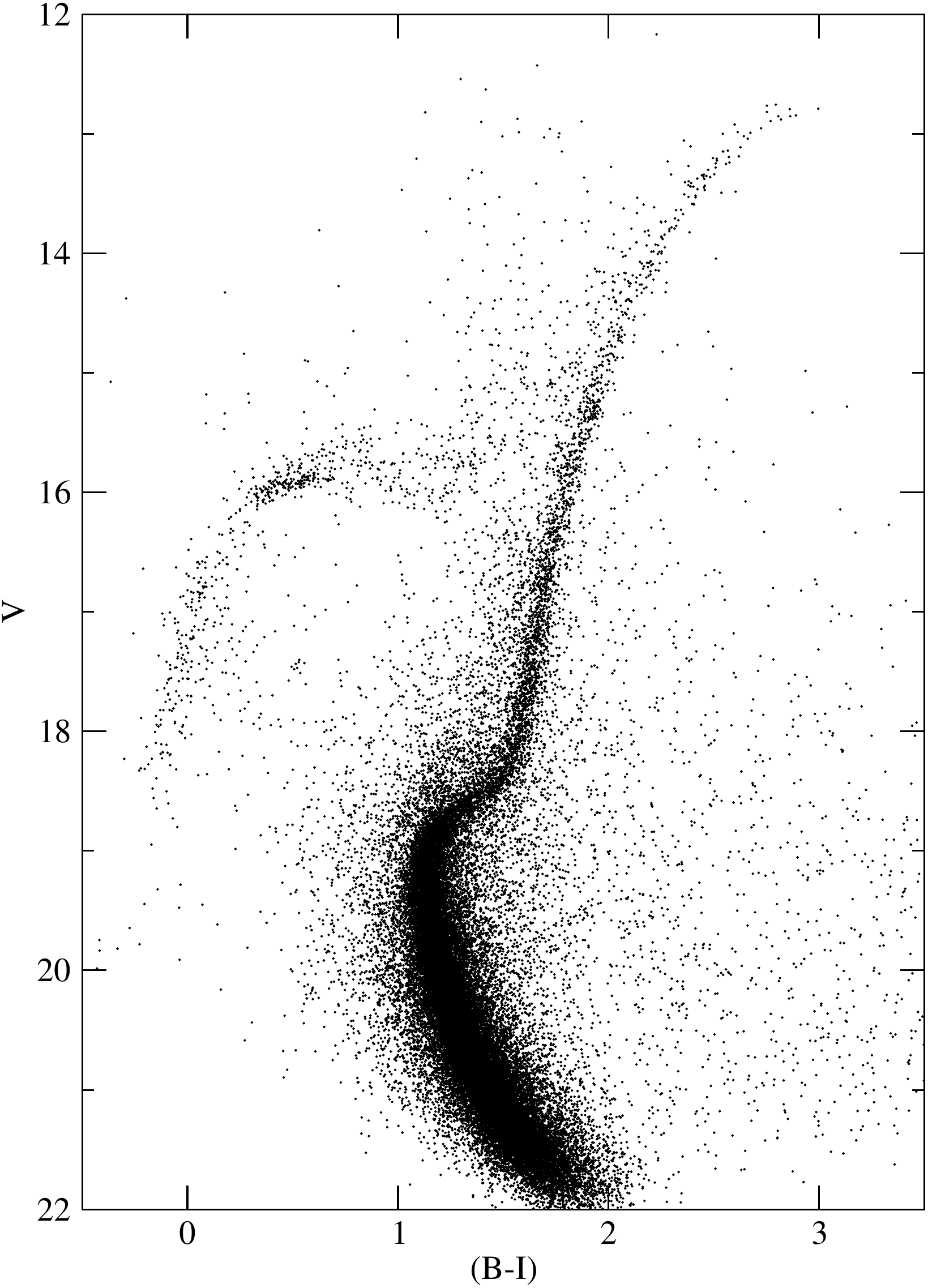}
\caption[V vs (B-I) Color-Magnitude Diagram]{The $V$ vs $(B-I)$ color magnitude diagram of M15. This CMD contains over 44,000 stars and extends more than three magnitudes below the MSTO.  This color and magnitude choice ensures that errors in color and magnitude are uncorrelated.}
\label{fig:bvicmd}
\end{figure}

In total, 44,800 stars were included in the final photometry file.  This compares well to a similar study of the globular cluster M92 measured 34,242 stars from the ground-based data \citep{paust}.  The main sequence of our photometry extends several magnitudes below the MSTO, as can be seen in Figure \ref{fig:bvicmd}.  This is a significant improvement over the previous work of \citet{ruelas}.  Our $BVI$ photometry can be found in Table \ref{photometry}.

\begin{deluxetable}{rccrrrrrr}
\tablewidth{0pt}
\tablecaption{$BVI$ Photometry of M15 \label{photometry}}
\tablehead{
\colhead{Num.} & \colhead{RA} & \colhead{DEC} & \colhead{$B$} & \colhead{$\sigma_B$} & \colhead{$V$} & \colhead{$\sigma_V$} & \colhead{$I$} & \colhead{$\sigma_I$}}

\startdata
5000115 & 21:31:03.95 & 12:06:28.9 & 19.837 & 0.047 & 19.182 & 0.036 & 18.620 & 0.063 \\
5000225 & 21:30:57.85 & 12:18:54.6 & 20.486 & 0.063 & 19.798 & 0.053 & 18.923 & 0.058 \\
5000246 & 21:30:56.50 & 12:18:53.3 & 18.700 & 0.013 & 17.596 & 0.014 & 16.627 & 0.023 \\
5000263 & 21:30:55.61 & 12:17:58.4 & 19.236 & 0.013 & 18.728 & 0.017 & 18.363 & 0.032 \\
5000290 & 21:30:53.81 & 12:10:27.4 & 18.926 & 0.015 & 18.135 & 0.014 & 17.522 & 0.018 \\
5050939 & 21:29:01.89 & 12:03:01.9 & 20.215 & 0.030 & 18.739 & 0.022 & 16.914 & 0.017 \\
5000299 & 21:30:52.93 & 12:15:59.7 & 18.209 & 0.010 & 17.327 & 0.011 & 16.679 & 0.009 \\
5000304 & 21:30:52.55 & 12:05:16.6 & 16.468 & 0.007 & 15.648 & 0.016 & 14.978 & 0.013 \\
5000308 & 21:30:52.43 & 12:03:30.2 & 19.166 & 0.048 & 18.156 & 0.029 & 16.850 & 0.025 \\
5000309 & 21:30:52.42 & 12:03:29.2 & 16.942 & 0.012 & 16.189 & 0.008 & 15.646 & 0.020 \\
5000311 & 21:30:52.50 & 12:15:47.6 & 16.187 & 0.005 & 14.779 & 0.004 & 13.691 & 0.006 \\
5000320 & 21:30:52.04 & 12:08:14.0 & 21.033 & 0.023 & 19.785 & 0.066 & 18.551 & 0.039 \\
5000327 & 21:30:51.88 & 12:06:02.6 & 22.078 & 0.042 & 21.401 & 0.043 & 20.396 & 0.048 \\
\enddata
\tablecomments{Table 1 is published in its entirety in the electronic edition.  The portion shown here is for guidance regarding form and content.}
\end{deluxetable}

\begin{figure}
\epsscale{0.7}
\plotone{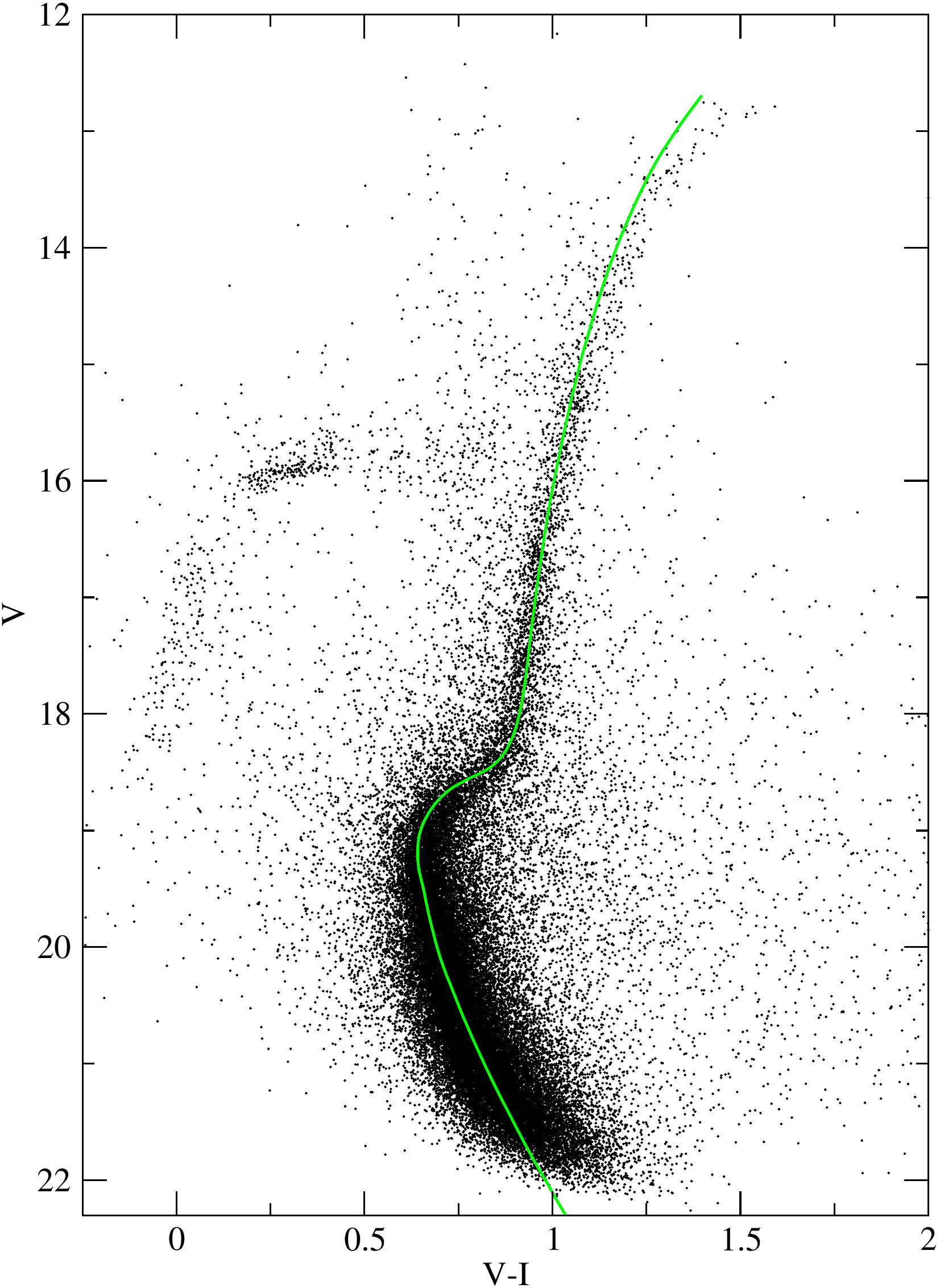}
\caption[V vs (V-I) Color-Magnitude Diagram with Isochrone]{Isochrone fit to the $V$ vs $(V-I)$ color magnitude diagram. The best fit isochrone represents a stellar population with an age of 13.0 Gyr, a distance modulus of $(m-M)_V = 15.4$, and abundances of $[\mathrm{Fe/H}] = -2.10$ and [$\alpha$/Fe] $= 0.2$. The age is uncertain to 1 Gyr and the distance modulus is uncertain at the 0.1 mag level.}
\label{fig:vicmdiso}
\end{figure}

Isochrone fitting of our photometry utilized DSEP \citep{dotter2007} models.    After extensive testing, values from the literature provided the best fits to our data.  Comparing the slope of the observed and isochrone RGBs, we determined the cluster metallicity to be [Fe/H]=$-2.10$.  This is very close to the value found by  \citet{zinn} and \citet{carretta} of [Fe/H]$=-2.15$ .  \citep{kirby} measured the [$\alpha$/Fe]=0.2 and we find this abundance gives the best match with our data.  Using these abundances, we find an age of $13.0 \pm 1.0$ Gyr, a distance modulus of $15.4 \pm 0.1$, and a color excess of $\mathrm{E}(V-I)=0.14 \pm0.02$.  In order to simultaneously determine these quantities, the color of the base of the RGB was used to determine reddening while the color and magnitude of the MSTO were used to determine age and distance.  The uncertainties quoted here are derived exclusively from the fitting.  They do not include systematic uncertainties, such as the age-reddening correlation.  With that limitation, the uncertainties are very robust. The isochrone fit to our $V$ versus $(V-I)$ CMD is shown in Figure \ref{fig:vicmdiso}.  The unreddened distance modulus is calculated using results from \citet{schlafly2011}:  $\mathrm{E}(B-V) = 1.25\ \mathrm{E}(V-I)$.  Using a typical value for extinction in the $V$ band, $\mathrm{A}_V = 3.1\ \mathrm{E}(B-V)$, our color excess results in an extinction of $0.54 \pm 0.08$ magnitudes in $V$.  This results in a dereddened distance modulus of 14.86, consistent with previous results \citep{majaess}.

\section{Luminosity Functions \label{lfs}}

\subsection{Completeness}

Extensive artificial star tests were performed to accurately determine the completeness of the photometry.  These tests were accomplished by adding artificial stars to the images utilizing the known PSF and reprocessing the images using our photometry pipeline.  The fraction of the artificial stars recovered, as a function of magnitude and stellar crowding, is assumed to be the completeness of the real star photometry at the same magnitude and crowding, albeit with small biases since the artificial stars are input and recovered with the same PSF and the real stars have more variable PSFs.  This bias is minimized by carefully characterizing the PSF.  Radius from the cluster center was used as a proxy for stellar crowding.  The uncertainty in the completeness is governed by Poisson statistics, therefore, large numbers of artificial stars are needed.  However, adding a large number of artificial stars to any individual image will change the crowding and therefore the completeness. To avoid this problem, many runs through the artificial star test are used with relatively small numbers of stars added in each run. 

For this work, a master list of artificial stars was generated with a $B$ magnitude and random R.A. and decl. positions.  We choose the $B$ magnitudes with a distribution biased towards fainter stars.  Since the fainter stars are always less complete, this maintains nearly a constant number of recovered stars as a function of magnitude.  $V$ and $I$ magnitudes were generated from the $B$ magnitude using the observed colors of the cluster along the RGB, AGB, and MS.  The images containing artificial stars were then measured using the same DAOPHOT and ALLSTAR pipeline as the initial photometry with a final matching step to compare the input artificial star list with the photometry. The recovered magnitudes were required to be within 0.1 mag of the input magnitudes to remove any possible confusion between real stars or blends of artificial stars.  In all, 370,000 stars were input and 177,142 were recovered.  The relatively small fraction recovered is a result of having the magnitude range of the artificial stars extend beyond the photometric detection limit.   The artificial star results were then used to create a completeness look-up table, as a function of cluster radius and magnitude, and this table was used to assign a completeness fraction to each individual real star.   The central arcminute of the cluster, where crowding is the most severe, has a completeness limit equal to the magnitude of the MSTO.  To ensure that the entire sample has a uniform photometric depth, this central region was removed from the photometry and analysis.

\begin{figure}
\epsscale{1.0}
\plotone{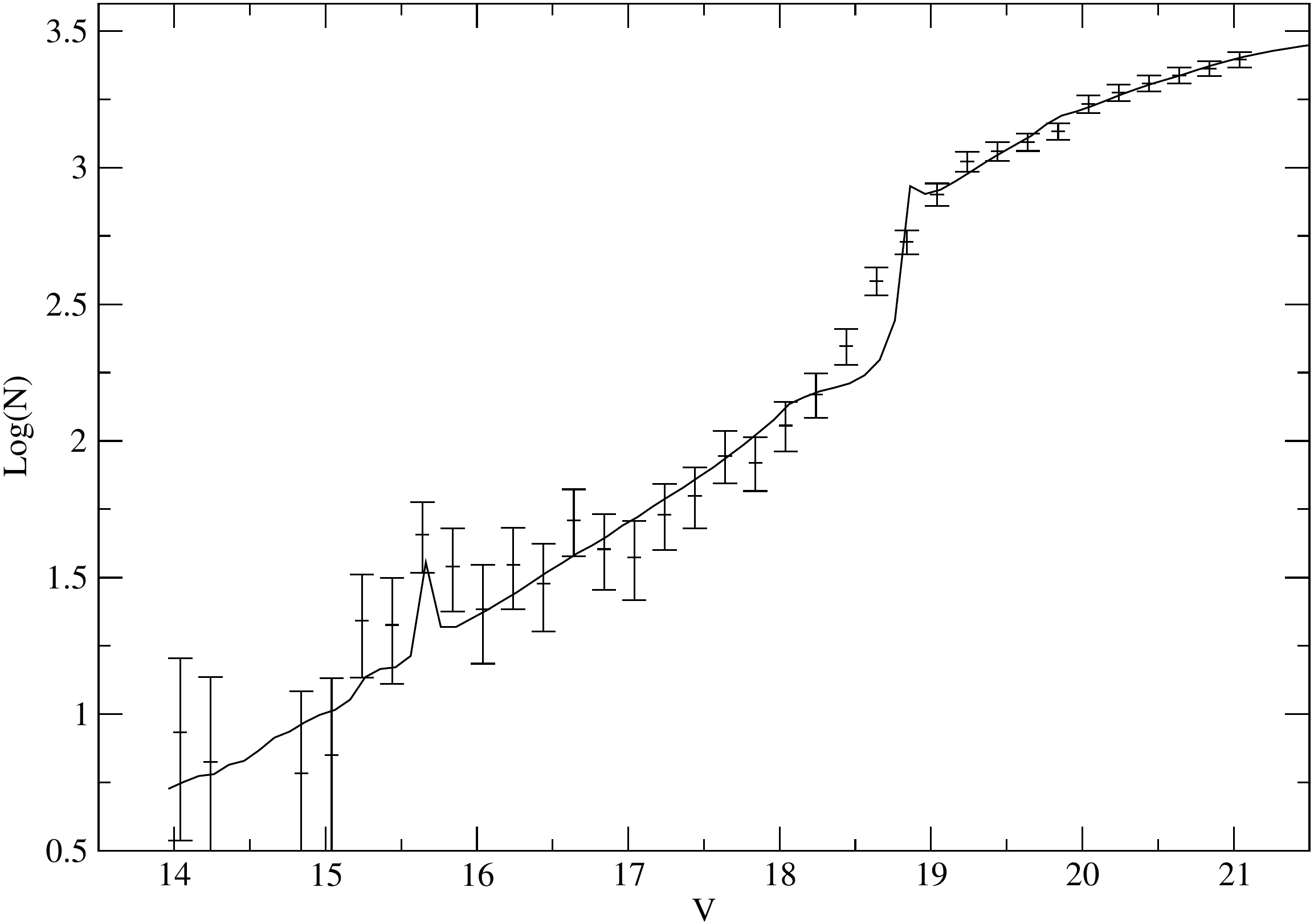}
\caption[V Luminosity Function]{The $V$ luminosity function fitted with DSEP models. The model shows has a Salpeter slope ($\alpha=-2.35$) and matches the best-fit isochrone.  The error bars shown are $2\sigma$ Poisson errors.  The main sequence does not cover a large enough mass range to be sensitive to the mass function slope.}
\label{fig:vlf}
\end{figure}

\subsection{$V$ Luminosity Function}

Before binning the photometry into a LF, the data was cleaned to ensure that only cluster stars on the RGB, SGB, and MS are represented. The asymptotic giant branch (AGB) stars and horizontal branch (HB) stars were removed using a color cut, a line was fit just blue-ward of  the RGB and MSTO and stars redder than the line were retained for the stellar sample.  Field contamination was addressed by removing all stars further than $3\sigma_{(V-I)}$ from the cluster ridge line.  This leaves some field stars with colors and magnitudes that match the cluster colors and magnitudes by coincidence, however, these stars are greatly outnumbered by cluster stars.  After these cuts were made in color-magnitude space, the LF was computed by summing the weight of each star in a given magnitude bin.  The weight is equal to $1/\mathrm{completeness}$.

Table \ref{tab:vlf} presents the $V$ LF and Figure \ref{fig:vlf} shows the observed LF with a theoretical LF matching our best fit isochrone.    The theoretical LF shown was generated with an $\alpha=-2.35$ mass function slope.  Due to the relatively small mass range seen on the MS, our fit is insensitive to the true slope of the MF.  Overall, the theoretical and observed LFs show good agreement along the MS and RGB.  Furthermore, the theoretical LF shows the correct position of the RGB bump.  The RGB bump is caused by a pause in stellar evolution along the RGB as the outward-moving shell burning hits material rich in H.  The position of this bump is highly sensitive to the position of the convective zone in the star.  The accurate bump placement suggests that DSEP correctly handles convection in old metal-poor stars.  While \citet{bevington} have suggested that the LF is not a good age indicator, our match between CMD and LF parameters suggests that the LF is an acceptable age indicator, supporting the work of \citet{paust}. 

\begin{deluxetable}{crrcccc}
\tablewidth{0pt}
\tablecaption{The M15 $V$ LF \label{tab:vlf}}
\tablehead{
\colhead{$V$} & \colhead{N} & \colhead{CN} & \colhead{Comp.} & \colhead{$\log(\mathrm{CN})$} & \colhead{+2-$\sigma$} & \colhead{$-2$-$\sigma$}}
\startdata
14.04 & 8 & 8.58 & 0.93 & 0.93 & 1.17 & 0.40\\
14.24 & 6 & 6.69 & 0.90 & 0.83 & 1.08 & 0.09\\
14.64 & \nodata & \nodata & \nodata & \nodata & \nodata & \nodata \\
14.84 & 6 & 6.08 & 0.99 & 0.78 & 1.04 & 0.05\\
15.04 & 7 & 7.10 & 0.99 & 0.85 & 1.10 & 0.24\\
15.24 & 22 & 22.02 & 1.00 & 1.34 & 1.50 & 1.10\\
15.44 & 21 & 21.23 & 0.99 & 1.33 & 1.48 & 1.08\\
15.64 & 44 & 45.36 & 0.97 & 1.66 & 1.77 & 1.50\\
15.84 & 34 & 34.73 & 0.98 & 1.54 & 1.67 & 1.36\\
16.04 & 24 & 24.25 & 0.99 & 1.38 & 1.53 & 1.16\\
16.24 & 35 & 35.16 & 1.00 & 1.55 & 1.67 & 1.37\\
16.44 & 30 & 30.10 & 1.00 & 1.48 & 1.61 & 1.28\\
16.64 & 51 & 51.17 & 1.00 & 1.71 & 1.82 & 1.57\\
16.84 & 40 & 40.20 & 1.00 & 1.60 & 1.72 & 1.44\\
17.04 & 37 & 37.47 & 0.99 & 1.57 & 1.70 & 1.40\\
17.24 & 53 & 53.68 & 0.99 & 1.73 & 1.84 & 1.59\\
17.44 & 62 & 62.97 & 0.99 & 1.80 & 1.90 & 1.67\\
17.64 & 86 & 88.24 & 0.97 & 1.95 & 2.03 & 1.84\\
17.84 & 81 & 83.21 & 0.97 & 1.92 & 2.01 & 1.81\\
18.04 & 103 & 113.88 & 0.90 & 2.06 & 2.13 & 1.96\\
18.24 & 131 & 147.80 & 0.89 & 2.17 & 2.24 & 2.09\\
18.44 & 198 & 222.31 & 0.89 & 2.35 & 2.40 & 2.28\\
18.64 & 334 & 384.46 & 0.87 & 2.58 & 2.63 & 2.53\\
18.84 & 465 & 533.91 & 0.87 & 2.73 & 2.77 & 2.69\\
19.04 & 603 & 797.87 & 0.77 & 2.90 & 2.94 & 2.87\\
19.24 & 770 & 1052.45 & 0.73 & 3.02 & 3.05 & 2.99\\
19.44 & 850 & 1146.40 & 0.74 & 3.06 & 3.09 & 3.03\\
19.64 & 962 & 1238.10 & 0.78 & 3.09 & 3.12 & 3.06\\
19.84 & 1066 & 1355.36 & 0.79 & 3.13 & 3.16 & 3.10\\
20.04 & 1131 & 1708.81 & 0.66 & 3.23 & 3.26 & 3.21\\
20.24 & 1237 & 1879.77 & 0.66 & 3.27 & 3.30 & 3.25\\
20.44 & 1329 & 2033.87 & 0.65 & 3.31 & 3.33 & 3.28\\
20.64 & 1409 & 2174.84 & 0.65 & 3.34 & 3.36 & 3.31\\
20.84 & 1571 & 2304.05 & 0.68 & 3.36 & 3.38 & 3.34\\
21.04 & 1534 & 2480.23 & 0.62 & 3.39 & 3.41 & 3.37\\
\enddata
\tablecomments{N is the raw number of stars measured per bin.  CN is the number of stars per bin corrected for completeness.}
\end{deluxetable}

\section{Conclusion \label{conclusion}}

1. We present high-quality deep photometry of the GGC M15 in the $B$, $V$, and $I$ filters. This allows for the creation of cluster CMDs using a variety of colors, including the particularly useful $V$ vs $(B-I)$ diagram, which has uncorrelated magnitude and color uncertainties.  The CMDs extend approximately three magnitudes below the MSTO and show a clear AGB, HB, RGB, SGB, and MS sequences.

2. Using DSEP models, isochrone fitting results in an age of $13.0 \pm 1.0$ Gyr, a distance modulus of $15.3 \pm 0.1$, a metallicity of $-2.10 \pm 0.1$ with an $\alpha$-enhancement of $0.2$. After dereddening, the distance modulus and age agree with results from recent literature.

3. Using extensive artificial star tests, we present a completeness-corrected LF in V-band.  The theoretical LF from DSEP models agrees with the observed LF, matching the stellar counts along both the RGB and MS and correctly predicting the position of the RGB bump.  This supports the results from \citet{paust} suggesting that DSEP models have overcome the problems found earlier by \citet{bolte} and \citet{sandquist}.

\acknowledgments
Facilities: \facility{Hiltner}

\end{document}